# Light multi-GEM detector for high-resolution tracking systems.

A.Bondar [a], A.Buzulutskov [a], R.de Oliveira [b], L.Ropelewski [b], F.Sauli [b] and L.Shekhtman [a] [*]

[a] *Budker Institute of Nuclear Physics, Novosibirsk, Russian Federation*
[b] *CERN, Geneva, Switzerland*

**Abstract**

Controlled etching of copper electrodes in Gas Electron Multiplier foils allows a reduction of the material budget by more than a factor of two for a triple-GEM detector. Detectors making use of thinned foils provide performances similar to those obtained with standard devices: a gain above $10^4$ for a double-GEM, with energy resolution of 27 % fwhm for 5.9 keV X-rays.

*Keywords:* GEM, high-resolution tracking, B-factory detectors

___________________________________________________________________

Readout systems for Time Projection Chambers (TPC) based on wire chambers, as well as tracking drift chambers used in current experiments or planned at high luminosity electron-positron colliders suffer from limited rate capability. The high occupancy in systems having large size resolution elements (close to a cm in most devices) limits the track finding capability and degrades momentum resolution.

Detectors based on the Gas Electron Multipliers (GEM)[1] can offer a solution, and allow overcoming the problem of limited rate capability and of high occupancy. Using several cascaded GEMs [2] permits to share the avalanche amplification between the stages, thus minimizing the probability of discharge [3, 4].

A medium-size system of GEM detectors, constructed for the high rate COMPASS experiment at CERN, is in operation since 2002 [5]; GEM detectors have also been adopted for the LHCb muon trigger [6], for the TOTEM experiment at CERN [7], and for PHENIX at BNL [8]. In each of these projects, sets of triple-GEM detectors cover a total active area of more than 1 square meter.

The amount of material in a tracking system is an important issue, as it can degrade the momentum resolution due to multiple Coulomb scattering, especially for low momentum particles. In a triple-GEM detector, such as those built for COMPASS, the total material thicknesses is about $3.4 \times 10^{-3}$ of a radiation length. To this, the copper layers of the standard GEM and drift electrodes (7 times 5 µm) contribute to about 70%. This is slightly more than for a 300 µm thick silicon microstrip detector (~$3.2 \times 10^{-3}$ $X_0$). However, in the case of GEM-based detector there is a possibility to substantially decrease the amount of material reducing the thickness of the copper layers; going from 5 to 1 µm, the total radiation length is reduced to $1.5 \times 10^{-3}$, about one half of a solid state device. More savings in the material budget could be achieved using a double-GEM assembly and a gas mixture based on light noble gases like Ne or He.

In this note we demonstrate the possibility to manufacture GEM foils with thin copper layer, and show that the main operational characteristics of a thin-GEM based detector, such as gain and energy resolution, do not differ from those of standard devices.

To manufacture foils with thin copper layers we started with standard, 5 µm Cu-coated GEMs; after cleaning and removal of the protective copper chromate layer with HCl, the foils were dipped in ammonium persulfate to do a micro-etch. During the process, the total thickness was monitored with a micrometer. When the desired thickness was reached, the foils were removed from the etching bath, washed and processed with chromic acid to form again the protective chromate layer.

The uniformity of thickness of the etched copper layer was measured on a sample foil 5x30 cm in size, undergoing the same etching procedure; along the 30 cm side the copper layer was found to have an average thickness of 1 µm, with about 10% uniformity.

Examples of Scanning Electron Microscope images of the holes for regular and etched foils are shown in Fig. 1; in the etched GEM, the copper layer is also removed around the hole by about four µm. A possible benefit of this geometry is to prevent the presence of sharp metal edges around the rim of the holes, with a consequent better protection against discharges.

For the experimental verification of performances, the copper electrodes of standard GEM foils, with 10x10 cm$^2$ active area, were reduced to 1 µm as described and assembled in a double-GEM detector with 2 mm transfer gaps; a drift electrode of aluminized mylar was added 3 mm above the first GEM. The detector was irradiated with an X-ray tube with Mo anode operating at 12 kV, or with an $Fe^{55}$ source. Proper choice of the applied potentials permitted to measure the performances of the detector both in the single and a double-GEM configuration. Equal value resistors (5 MΩ) were used in the chain for powering the detector, so that all gaps and GEM foils were fed with the same potential difference. As filling gas, we used a 70-30 mixture of argon and $CO_2$.

The results of the gain measurements for thin single- and double-GEM assemblies are shown in Fig. 2; error bars include fluctuations due to temperature and pressure variations. For comparison, results obtained with a standard double-GEM are also plotted in the figure (from Ref. 4): it can be seen that the gain performances are substantially identical. The pulse height spectrum for 5.9 keV X-rays from $Fe^{55}$ source is shown in Fig. 3.

The presence of insulating rims around the light GEM holes suggests that these foils might show a larger gain instability than regular GEM. The gain stability after application of high voltage was measured with a single GEM detector, using either standard or light foils, and recording the anode current under irradiation. During the measurement the detector was kept continuously under voltage, but exposed to irradiation only for a short periods (several seconds) needed measure the gain. The results of the comparison are shown in Fig.4; for each GEM type two sets of curves are shown for different induction fields. Error bars represent systematic variations due of ambient pressure and temperature, as well as fluctuations of the current readings due to pick-up noise. In all cases the gain shows an initial increase during the first hours of operation; this increase is more pronounced for lower induction field and is about 20% for the induction field of 2.5kV for both regular and light GEMs This is compatible with previous observations [9].

This is the first demonstration that GEM foils with thin copper layer can be easily produced and shows performances similar to the standard, thicker foils. The feasibility of the mechanical design of low-mass detectors, using thin foils and including self-supporting cylindrical GEM structures is also under investigation at CERN and at BINP.

## Figure captions

Fig. 1: SEM image of a standard (left) and light GEM hole (right).
Fig. 2: Effective gain as a function of GEM voltage for Single- and Double-GEM set-ups. Data for regular Double-GEM from [4], are shown for comparison.
Fig. 3: Pulse height spectrum measured with 5.9 keV X-rays with light Double-GEM at a gain of ~50000.
Fig. 4: Gain variation after application of high voltage for light and regular GEMs, for several value of transfer field.

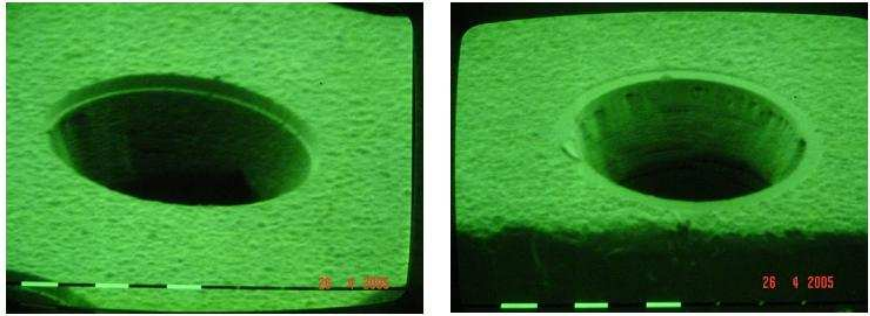

Fig.1.

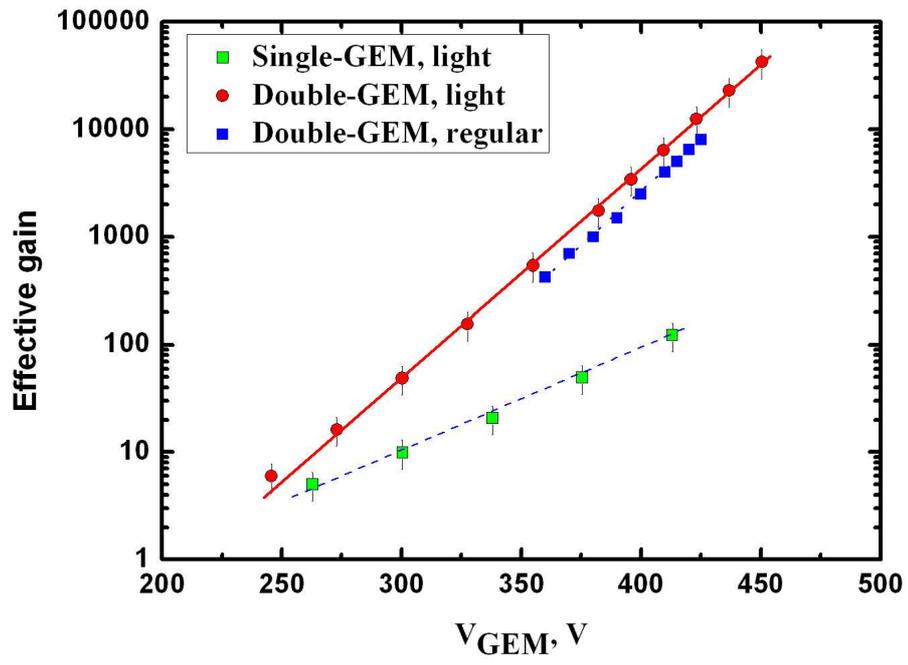

Fig.2.

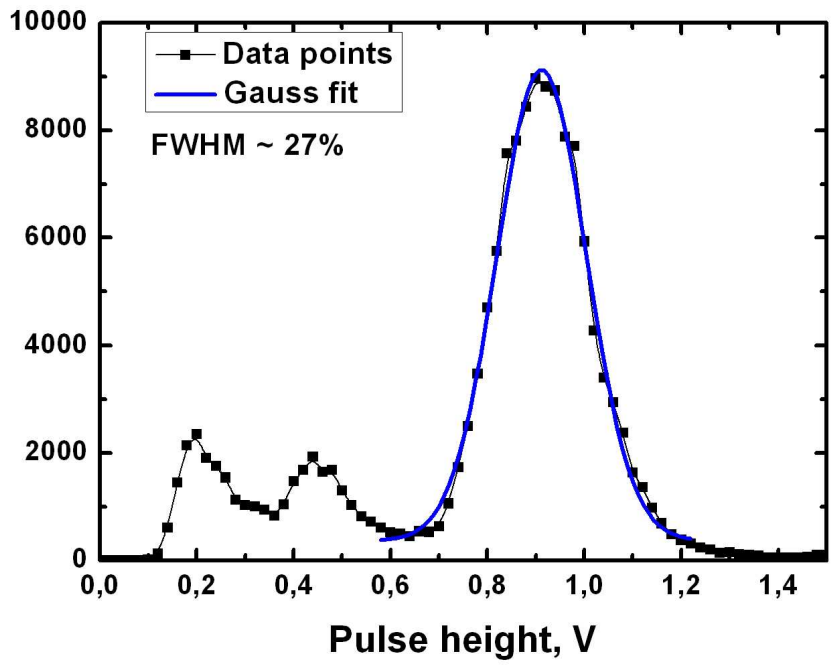

Fig.3.

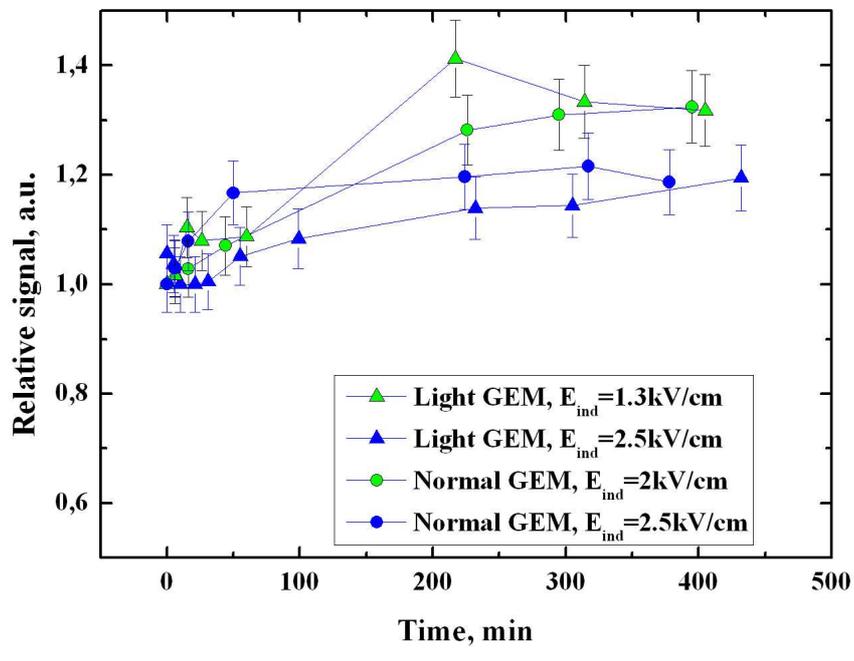

Fig.4.